\documentclass[sigconf]{acmart}

\AtBeginDocument{%
  \providecommand\BibTeX{{%
    \normalfont B\kern-0.5em{\scshape i\kern-0.25em b}\kern-0.8em\TeX}}}

\copyrightyear{2024}
\acmYear{2024}
\setcopyright{rightsretained}
\acmConference[ITiCSE 2024]{Proceedings of the 2024 Innovation and
Technology in Computer Science Education V. 1}{July 8--10, 2024}{Milan, Italy}
\acmBooktitle{Proceedings of the 2024 Innovation and Technology in Computer
Science Education V. 1 (ITiCSE 2024), July 8--10, 2024, Milan,
Italy}
\acmDOI{10.1145/3649217.3653624}
\acmISBN{979-8-4007-0600-4/24/07}

\usepackage{comment}
\usepackage[inline]{enumitem} 
\usepackage{listings}
\usepackage{multirow}
\usepackage{placeins}
\usepackage{colortbl}
\usepackage{multicol}
\usepackage{stfloats}
\usepackage{dirtytalk}
\usepackage{caption} 
\usepackage{balance}
\usepackage{tabularx}
\usepackage{array}
\usepackage{pifont}

\hypersetup{
           breaklinks=true,   
           colorlinks=true,   
           pdfusetitle=true,  
        }

\usepackage{breakurl}

  {\list{}{\leftmargin=0.2in\rightmargin=0.2in}\item[]}%
  {\endlist}

\newcolumntype{s}{>{\hsize=.06\hsize}X} 
\begin{document}
\title{Student Perspectives on Using a Large Language Model (LLM) for an Assignment on Professional Ethics}



 \author{Virginia Grande}
   \email{virginia.grande@it.uu.se}
\affiliation{%
 \institution{Uppsala University}
 \streetaddress{}
 \city{Uppsala}
 \country{Sweden}}

\author{Natalie Kiesler}
\email{natalie.kiesler@th-nuernberg.de}
\orcid{0000-0002-6843-2729}
\affiliation{%
 \institution{Nuremberg Tech}
 \streetaddress{Keßlerplatz 12}
 \postcode{90489}
 \city{Nuremberg}
 \country{Germany}}

\author{Mar\'ia~Andre\'ina Francisco~R.}
\email{maria.andreina.francisco@it.uu.se}
\orcid{0000-0001-8745-9858}
\affiliation{%
 \institution{Uppsala University}
 \streetaddress{}
 \city{Uppsala}
 \country{Sweden}}





\begin{abstract}
The advent of Large Language Models (LLMs) started a serious discussion among educators on how LLMs would affect, e.g., curricula, assessments, and students' competencies. Generative AI and LLMs also raised ethical questions and concerns for computing educators and professionals. 

This experience report presents an assignment within a course on professional competencies, including some related to ethics, that computing master's students need in their careers. For the assignment, student groups discussed the ethical process by Lennerfors et al. by analyzing a case: a fictional researcher considers whether to attend the real CHI 2024 conference in Hawaii. The tasks were (1) to participate in in-class discussions on the case, (2) to use an LLM of their choice as a discussion partner for said case, and (3) to document both discussions, reflecting on their use of the LLM. 

Students reported positive experiences with the LLM as a way to increase their knowledge and understanding, although some identified limitations. The LLM provided a wider set of options for action in the studied case, including unfeasible ones. The LLM would not select a course of action, so students had to choose themselves, which they saw as coherent.

From the educators' perspective, there is a need for more instruction for students using LLMs: some students did not perceive the tools as such but rather as an authoritative knowledge base. Therefore, this work has implications for educators considering the use of LLMs as discussion partners or tools to practice critical thinking, especially in computing ethics education.

\end{abstract}


\maketitle

\section{Introduction}

Students pursuing a master's degree in a computing area in Sweden must acquire a set of professional competencies. In particular, the Swedish Higher Education Ordinance specifies that students who complete a master's program should have demonstrated:
\begin{quote}
    \textit{the ability to make assessments in the main field of study informed by relevant disciplinary, social and ethical issues and also to demonstrate awareness of ethical aspects of research and development work.} \cite{highereducationordinance}.
\end{quote}


Some students have experience in ethics in computing and/or academic writing before they start their master's, e.g., a bachelor's degree from an institution that requires a bachelor's thesis. However, students' backgrounds vary when considering an international cohort. Thus, there is a need for a course that aims to address and harmonize the development of students' professional competencies.

Generative Artificial Intelligence (AI), and Large Language Models (LLMs) in particular, offer a new avenue for students to discuss and report their conclusions and ideas, not only in general, but also related to social and ethical issues. In this experience report, we describe how students experienced using an LLM as part of discussing and writing a group assignment on professional ethics. An LLM was used not only for improving and commenting on academic writing but also as a discussion partner to provide explanations on ethical theories and how to apply them to a fictional scenario. In the following, we introduce the course design and the theory in engineering ethics education that the course assignment was based on, and categorize the students' experiences before concluding and presenting the lessons learned from this assignment design.

\section{The Role of Generative AI in Computing Education}
 
Ever since Large Language Models and related applications (e.g., Codex, ChatGPT, Copilot) were released to a greater public, their implications on education have become subject to intensive research. 
A 2023 ITiCSE working group report~\cite{prather2023robots} implies that computing educators should focus on exploring and navigating the generative AI revolution instead of ignoring recent developments. They reviewed and classified recent literature into five categories:
\begin{enumerate} 
    \item Assessing the performance and limitations of LLMs related to the generation of code as a solution to tasks~\cite{finnieansley2022the,finnieansley2023my,kiesler2023large}, multiple-choice or exam questions~\cite{savelka2023can,savelka2023large,dobslaw2023experiences}. 
    \item Position papers, surveys or interviews addressing implications~\cite{becker2023programming,becker2023generative,macneil2024bof} for education (e.g., curricula, pedagogy, assessment, etc.).
    \item Using LLMs to analyze students' solutions and provide feedback~\cite{phung2023generating,kiesler2023exploring,azaiz2024feedbackgeneration}, or fix errors~\cite{ahmed2022synshine,zhang2022repairing,roest2023nextstep}. 
    \item Using LLMs to create teaching material (e.g., programming exercises~\cite{denny2022robosourcing}, multiple-choice questions~\cite{tran2023generating}) or learning objectives~\cite{sridhar2023harnessing}.
    \item Studying the interactions between programmers (including students) and LLMs w.r.t. interaction patterns~\cite{prather2023its,vaithilingam2022expectation,barke2023grounded}, productivity~\cite{kazemitabaar2023studying,nam2023in}, code security~\cite{sandoval2023lost}, and students' use of code explanations~\cite{macneil2023experiences,pankiewicz2023large}. 
\end{enumerate}

This experience report differs from prior work, focusing on a pedagogical design of students interacting with LLMs (in this case, ChatGPT). This paper reflects on the application of ChatGPT in a classroom scenario of a master's program course module on ethics in computing, where students learn the ethical process according to Lennerfors et al.~\cite{LennerforsFIE20}. The application context is thus not the generation of code or feedback (e.g.,~\cite{kazemitabaar2024codeaid}), but the identification of ethical dilemmas, reflecting, and responding to them, which has not been addressed in prior studies or experience reports.

\section{Course Design}

A Swedish university has an introductory course for students who are starting their master's program in one of two computing areas: computer science or embedded systems. Administratively, these are two separate courses of 5~ECTS (European Credit Transfer and Accumulation System) credits each. In practice, they are given as a joint course over the course of two months in the fall semester. In this paper, ``the course" thus refers to the joint course with students from both programs. As with other education at the graduate level in this university, the language of instruction is English.


The goal of the course is for students to develop a set of professional competencies needed to succeed in their master's-level studies and potential professional careers in the particular context of Sweden. Thus, international students from several continents enroll and the syllabus is very diverse, touching upon many different topics, including:
\begin{itemize}
    \item Preparation for defending a master's thesis: academic writing (with \LaTeX) and presentations,
    \item Ethics in computing,
    \item Equal opportunities (in other contexts known as DEIA: Diversity, Equity, Inclusion, and Access) in computing and in the Swedish context/legislation,
    \item Academic honesty,
    \item Introduction to C programming,
    \item Project work with Raspberry Pi, etc.
\end{itemize}

Different assignments throughout the course assess these various topics. As the course grade is pass or fail, all assignments are only graded as pass or fail, and all assignments (with opportunities for re-submission after feedback) must be passed to pass the course. In this paper, we focus on an assignment aimed at evaluating competencies related to, mainly, ethics in computing and academic writing. However, other parts of the course are assessed too, e.g., providing a correct reference list and citations for academic honesty.

\subsection{Theoretical Underpinning: Ethics Education}

We used the framework that Lennerfors et al.~\cite{LennerforsFIE20} describe for teaching the ethical process. It consists of four aspects:
\begin{itemize}
    \item \textit{Awareness}: identifying that there is an ethical dilemma,
    \item \textit{Responsibility}: considering factors for ownership of the problem and identifying a problem owner,
    \item \textit{Critical thinking}: using a variety of ethical theories to reflect on the problem with diverse perspectives,
    \item \textit{Action}: one course of action is chosen from the set of potential ones discovered when applying critical thinking.
\end{itemize}

While these aspects are described linearly, they are a process where every aspect includes the aspect(s) before. The framework is explained at length in the textbook \cite{LennerforsBook19} we used for the course.

Lennerfors et al. identify three areas in terms of awareness:
\begin{itemize}
    \item Working with technology ``concerns the ethical issues that have to do with the development, refinement, implementation, or use of technology (physical or non-physical artefacts, including knowledge and skills to use the artefacts)." \cite[p.~2]{LennerforsFIE20}
    \item Working with others refers to relationships with individuals who are part of our professional lives, such as clients, colleagues, etc.
    \item Private life, which emphasizes ethical dilemmas about, for instance, work-life balance. This area is included to expand definitions of engineering ethics solely focused on the professional environment \cite{harris1996engineeringethics}.
\end{itemize}

This framework relies on the autonomy matrix~\cite{erlandsson2005autonomy} for scaffolding the aspect of critical thinking. The rows in the matrix are the different courses of action that the problem owner could take, and the columns represent the individual(s) affected by these actions. The matrix cells are the intersections where we reflect on how the particular individual(s) would be affected by the particular action. These effects are represented as pros (with a plus sign) or cons (with a minus sign) and a probability level (with H for High, M for Medium, and L for Low). Table~\ref{table:autonomyMatrixAndreas} shows an example of this matrix as a start for the assignment described in this paper.

Different ethical theories can be used to develop the autonomy matrix, as explained in the textbook~\cite{LennerforsBook19}. In the course, we chose several. Consequentialism is the theory that students may be quicker to understand, as they have experienced this approach of looking at who is affected by what actions. Virtue ethics, with a focus on developing desirable character traits (virtues), has been studied in computing education as a way of, e.g., developing professional values \cite{FrezzaVirtues22}. Care ethics (a feminist approach), focused on relationships instead of universal principles~\cite{gilligan1982different}, is needed in engineering~\cite{riley2013hiddenCare}  and in computing in particular~\cite{peters2020care}. Deontology, where the emphasis is on following universal rules, can be seen in engineering and computing in our agreement to follow the codes of ethics of professional organizations like ACM~\cite{ACMcodeconduct} and IEEE~\cite{ieee23codeethics}.

\subsection{The Group Assignment}
The course had two seminars on ethics in computing. During the first seminar, students were introduced to the awareness and responsibility part of the ethical process. They had in-class group discussions on these aspects using a provided fictional case based on a real situation. The same case and discussion format was used again in the second seminar after the aspects of critical thinking and action were presented together with different ethical theories. During both seminars, the students were asked to take notes of their discussions and the points that other groups made when the whole class summarized their conclusions.

At that point, a group assignment was introduced: students were to write a report on the case discussion. First, they would report their own conclusions based on their notes. After that, they would make use of an LLM of their choice as a discussion partner. The LLM was to be used to analyze the case using the ethical theories provided in a list. The students would report on the LLM's output, including the prompts used to elicit those answers, and comment on these from two angles: contributions to the analysis of the case, and their feedback on the academic writing level of the LLM.

\subsubsection{The case}

CHI, the major conference for Human-Computer Interaction (HCI), was announced to be held in Hawaii in 2024. This announcement sparked different reactions in the HCI community, with social media making these reactions more prominent to people within and outside the HCI community. Concerns were raised about the negative impact on, e.g., the environment due to flying being the main traveling option to attend the conference, and whether attendance would create problems that did not outweigh local jobs and tourism expenses~\cite{kiesler2024conferencesexclusive}. Some academics in HCI started, thus, to reconsider their participation in CHI 2024. This real situation was explained to the students. Then a fictional actor was presented for a case called ``Andreas' dilemma'':
\begin{quote}
Andreas, a fictional early-career academic, had a paper accepted to CHI 2024. After this acceptance, he learns about the concerns raised by the community and thus decides to reevaluate what to do. Andreas spends some time trying to increase his knowledge about the situation. He considers the arguments made about the negative impact and the ones he comes up with himself.
\end{quote}
Thus, Andreas has gathered the following perspectives:
\begin{itemize}
    \item This is the best conference to network and get contacts for future job opportunities:
    \begin{itemize}
        \item Andreas will soon be in the job market and needs to support his family.
        \item Andreas is fairly new in academia and lacks professional connections and networks.
    \end{itemize}
    \item None of the conference attendees are local to Hawaii:
        \begin{itemize}
        \item Everyone will fly there, as more sustainable travel options are infeasible.
    \end{itemize}
    \item There are no options to attend online.
    \item It is unclear whether the conference will generate (temporary) local jobs.
    \item CHI only takes place once a year, but there are other (smaller) conferences until the next CHI.
    \item If Andreas does not attend:
        \begin{itemize}
        \item His paper will not be published, which can damage his career.
        \item He will not have to worry about the time or nights away from his family.
    \end{itemize}
    \item Because of his financial situation, this may be Andreas’ only chance to visit Hawaii.
\end{itemize}

Table \ref{table:autonomyMatrixAndreas} shows a starting status of the autonomy matrix for critical thinking. The content shows what the teacher presented the students with, incorporating a few suggestions from the previous class discussion.

The last part of the assignment was for students to reflect on their experience using LLMs in this assignment, the course overall, and in their studies in general, if applicable. Here, we categorize the student experiences taken from these reflections and how they report their use of an LLM as a discussion partner in the findings of their report. While Swedish regulations reflect that submitted assignments are public documents, we still asked for the students' informed consent, and thus analyzed only the anonymized reports of 12 out of 16 groups, where each group had 4 members.

\begin{table*}[t!] 
    \centering
    \small
       \caption{Autonomy matrix as a start of discussing Andreas' case.}
    \label{table:autonomyMatrixAndreas}
   \begin{tabularx}{\textwidth}{@{}X | X | X | X |s|X@{}}
        \toprule
         & \multicolumn{5}{c}{\textbf{Affected}} \\
         \cmidrule(l){2-6}
        \textbf{Action} & Andreas himself & Andreas' family & Other academics & \dots{} & The planet \\ \midrule
        Flying to the conference & + Only chance to visit Hawaii (M) &  & + Knowledge exchange (H) &  & - Unsustainable travel impact (H)  \\ 
        Demand online attendance &  & + More time with family (H)  &  &  &  \\
        Withdraw the paper and submit it somewhere else & - Worse academic career prospects (M) &  &  &  &  \\
        Quit academia & + Higher salary (M) &  &  &  &  \\
        \dots{} &  &  &  &  &  \\
        \bottomrule
    \end{tabularx}
 
\end{table*}

\section{Experiences}\label{sec:results}
We analyzed the student texts to describe their experiences and how the teacher (first author) reflected as a result. This leads to lessons learned to be shared with the computing education research community.

While any LLM could be used for the assignment, all student groups chose to use a version of ChatGPT, and thus both terms appear in this section.

\subsection{Student Reflections}
Here we present different themes as results of our thematic coding analysis based on Braun and Clarke's~\cite{braun2006using}. Initial coding was influenced by the aspects of the ethical process~\cite{LennerforsFIE20} and later expanded with other topics in the data. The findings are presented with quotes where the student group order has been randomized from the original one, and each group has been assigned a character (from A to L).

\subsubsection{Increasing knowledge/understanding}
Students report using ChatGPT in their education to learn more about topics they need to understand further, as Group~L writes:
\begin{quote}
    \textit{ChatGPT is a helpful resource for me in my studies since it provides in-depth explanations and examples that help me better understand difficult concepts. [...] My educational experience has been greatly improved by using ChatGPT.}
\end{quote}

The appeal of customizing one's learning experience can be considered especially beneficial when time is of the essence, as Group~I points out:
\begin{quote}
    \textit{[LLMs] offer a personalized learning experience, adapting to our needs and thus making our study sessions more productive and enjoyable. [...] LLMs are adept at condensing lengthy texts into succinct summaries, making it easier for us to grasp the key points especially when we are pressed for time.}
\end{quote}

For the assignment, even though this was not instructed, a group used ChatGPT to understand better the real situation that the case referred to, as one would perhaps traditionally use a search engine. They used the prompt: ``Why are so many international conferences being held on remote islands or far away countries?''

\subsubsection{Limitations of support for understanding}

As opposed to the above, some students saw limitations to the support they could get from ChatGPT for the assignment and saw it more as a provider of definitions rather than a tool capable of an analytical process, as Group~L phrases it:
\begin{quote}
    \textit{The best example of that [ChatGPT's limitations] is how, after being prompted about the topic of Ethic Theories, it could only give textbook definitions without going in depth or broaching the situation analyzed [here].}
\end{quote}

Group~G highlights this further for ethics, where sticking to memorization is not enough for a contextualized ethical process following the aspects in the used framework \cite{LennerforsFIE20}: 
\begin{quote}
    \textit{[...] the LLM had simply adjusted its writing to match the basic ethical frameworks that were provided to it as a reference. This amounts to a sort of rote learning and application of ethics without any critical thinking or awareness. It is ethics in a void with no understand[ing] of context.}
\end{quote}

More generally, students were reluctant to fully trust the LLM output as it could ``be wrong and unreliable'' (Group~H). Group~E emphasizes the danger of not spotting incorrect output, something more likely to happen to those less knowledgeable in the area:
\begin{quote}
    \textit{One of the main dangers is that it can be hard to recognize that ChatGPT made a mistake if one is new to a topic and therefore cannot distinguish between a correct or false answer. As a result, it can be a dangerous road to solely believe the answers of ChatGPT and we highly recommend to validate everything that was taken from the platform and compare it with other sources as well.}
\end{quote}

ChatGPT also sometimes used ```sophisticated' (in reality, complicated) words that could be simplified" (Group~B), putting an unnecessary barrier to understanding. The above-praised increased speed of the learning process was seen as risky, as it could ``make us [students] lazy researchers if we rely on them too much'', so Group~D claims that they ``tried to use this technology carefully.'' Group~I agreed, pointing to concerns on academic dishonesty: ``As with all tools, careful and moderated use is imperative to ensure genuine learning and adherence to academic integrity.''

\subsubsection{Quality of support for critical thinking for the autonomy matrix}
When the students compared their in-class discussions with the LLM's output, they realized that it affected in several ways how they could develop the column for courses of action in the autonomy matrix. For instance, LLM suggested options that the students did not think of, e.g., that Andreas' colleagues could present for him (Group~F). 

A better understanding of ethical theories, thanks to the explanation of the LLM, could result in more options to add as courses of action. Group~E states that: 
\begin{quote}
    \textit{[ChatGPT] provided examples for the moral perspectives of each ethical theory and therefore allowed for a better understanding of the given ethical theory. As a result we learned how ethical theories prioritize different aspects and how they therefore come to different solutions.}
\end{quote}
Similarly, Group~L describes an example for virtue ethics:
\begin{quote}
    \textit{ChatGPT brought up that virtue ethics involves seeking guidance from others, point that we didn’t previously consider [...]. Andreas could have talked to his colleagues [...].}
\end{quote}

Infeasible options were suggested by ChatGPT, like online attendance (e.g., Groups~D and~F) despite this being clearly indicated in the description of the case; or biased ones, as Group~D reports that ChatGPT provided a ``[...] clear list of reasons and nuanced perspectives [...] somewhat positively biased [...] [ChatGPT] didn’t provide any negative implications." However, in some cases, the students saw some possibilities in using these erroneous suggestions, as Group~D reports: 
\begin{quote}
    \textit{Although this answer [that Andreas would face professional isolation if he withdraws his paper] might contradict previous answers or might be exaggerated [...], I can extrapolate some ideas that I can use successfully: that his decision can result in pushback in multiple groups.}
\end{quote}

Some students came up with factors that the LLM did not report, e.g., Group~K thought that if ``Andreas' career remains prolific, he might discover something within his field that is of great use to humanity.''

Perhaps this limitation was due to ChatGPT, as the students perceived it, sticking too strictly (at least initially) to what was described in the case description, as Group~G reflects: 
\begin{quote}
    \textit{[the results] presented by the LLM only arise from the description of [Andreas's] dilemma given as input, while we also came up with arguments by ourselves. Though, the LLM could probably also add further points, if we would request it.}
\end{quote}

\subsubsection{Deciding a course of action}
The students were not able to compare what course of action they chose as opposed to what the LLM chose, as ChatGPT was perceived as refraining from doing so. This was positively regarded by the students, perhaps enhanced by their technical background, as this group refers to the perspective of the developers and, even though this is not explicitly labeled as such, what can be interpreted as the developers' ethical responsibilities:
\begin{quote}
    \textit{ChatGPT responses always mention the ``two sides of the coin'', but never decide on one (which makes sense, as the developers of ChatGPT would not like for it to go making decisions on potentially delicate dilemmas).}
\end{quote}

Some students support this view, taking it one step further by highlighting the ``human side'' of ethics that they do not consider as fully reflected in the sole use of theory. Group~G claims that ``Ethical dilemmas for humans should be solved by humans as it concerns human character and morality far more than any ethical frame of reference.''
Group~K expands this thought by pointing out the limitations caused by the implementation of LLMs:
\begin{quote}
    \textit{[...] humanly touch is required as we having past experience we are in state of mind to take the decisions. LLMs can take decisions but they might change their opinions when we ask the same questions in repetition, this is because they are trained on a fixed set of data sets which can be outdated and lack the intelligence we as humans exhibit in [these] scenarios.}
\end{quote}

\subsubsection{Scaffolding that allows focus on critical thinking}
Overall, students seemed to have a fairly positive experience using an LLM as a discussion partner, even if they were aware of its limitations. The LLM was seen as enhancing the learning experience in many ways. Group~D expresses many examples:
\begin{quote}
    \textit{Not only that barriers such as language, academic expression rules, and lack of inspiration for simple concepts [are] now easier to overcome by the help of AI, but serious and tricky issues such as bias are also more manageable.} 
 \end{quote}
 \begin{quote}
    \textit{By applying well-practiced skills in LLM prompt engineering and maintaining a strong ethical foundation, a committed researcher can now explore a wider range of perspectives than they ever thought possible and also focus on experimenting, thinking, and analyzing rather than writing. We can even define this phenomenon as ``improved efficiency in meaningful research''.}
\end{quote}

LLMs are perceived to lift, for some, the weight of dealing with language barriers connected to language proficiency (where ChatGPT is used for grammar and spelling checking, as groups such as~E and~J did) or being unfamiliar with the formal structure of academic expressions. Time and energy previously spent on finding an appropriate way to communicate one's message are now left to the LLM; the person can focus instead on critical thinking.

While there is widely spread - and very relevant - worry in communities in computing and beyond about bias in LLMs, Group~D points in a different direction: that access to LLMs helps the user expand their worldviews, potentially mitigating their own biases.

Group~C refers to the benefit of a structure to frame ethical processes provided by the use of ChatGPT:
\begin{quote}
    \textit{[LLMs] provide a structured framework to assess complex ethical dilemmas, helping us consider various perspectives and weigh the potential impact of decisions. LLMs can be a useful tool in ethical discussions and decision-making processes in computing.}
\end{quote}

\subsection{Educator Perspective}
The assignment was designed for a focus on two out of four aspects of the ethical process: from awareness, responsibility, critical thinking and action, we defined the former two in the first ethics seminar, and the written assignment put more weight on the latter two. However, as all aspects are connected, it was possible to see references to all of them in explicit or implicit ways.

For awareness, for identifying an ethical dilemma, to the teacher LLMs did not seem like good support. In order to ask an LLM whether something is ethically problematic, one has to already have a suspicion that there may be a problem, or must ask specifically for every situation where there might be an ethical dilemma. However, awareness is connected to taking responsibility, which in turn connects to one's knowledge. As one group used ChatGPT to increase their knowledge of the case (e.g., why are international conferences held in remote places sometimes?), this could be an advantage of using LLMs in ethics education. Unfortunately, ChatGPT gave partially incorrect answers, which the students -not so familiar with the academic world - interpreted as completely correct. As the assignment asked to state the process followed with the LLM, it was possible to spot and address this issue.

Comparing the in-class discussions with the reports, the teacher perceived a general increase in understanding of the ethical theories and process. Thus, using LLMs enhanced critical thinking, expanding the students' views and knowledge. However, this was the part of the assignment were it was most noticeable how students with less confidence and/or experience in using LLMs would trust too much the output from ChatGPT.

The fact that ChatGPT did not provide a chosen course of action was another advantage of using an LLM, as the students had to make the choice themselves. The grading was not based on whether the teacher would have chosen the same course of action but rather on showing how that decision had been reached.

It is important to note that the assignment did not evaluate students' proficiency using an LLM but their reasoning when evaluating what to include in the autonomy matrix and why.

As the assignment is graded as pass or fail, with opportunities for resubmission until the end of the study period, the above-men\-tioned disadvantages could be addressed when giving feedback for each submission, such as clarifications of misconceptions, etc.

\section{Lessons Learned}

The assignment instructions referred to the LLM as a simulation of a discussion partner that can be challenged. However, this did not seem to be the perception of all students. Some students seemed to perceive ChatGPT more like an authoritative, knowledgeable figure whose knowledge is not questioned, as it is assumed to be correct. Therefore, more emphasis needs to be placed on the instruction that it is possible to critically analyze the output from the LLM and to actively encourage students to do so. 

Considering the possibility of falsified or hallucinated answers by the LLM, it may be worth considering the inclusion of more teaching staff during the writing of the report. This may take the form of a meeting to discuss how the analysis, in discussion with the LLM, is going, even though this was addressed after the submission.

Students appreciated that they could focus on their ideas instead of the language. Moreover, their proficiency in using the LLM use was not graded. However, different levels of it can still create disadvantages for some students. Knowing what prompts to use when can enrich the learning experience and lead to a deeper understanding of the case and the ethical theories. 

All of these aspects may be addressed by adapting the instruction to become more explicit, or by adding an introduction to LLMs, their characteristics, implementation, and resulting limitations. This would help set more equal conditions for all students, regardless of their prior knowledge or experiences with LLMs.

\section{Discussion}

Overall, the assignment worked well, and the students got additional positive experiences not possible in alternative formats. A conversation with a peer would not have provided the level of detailed explanations of ethical theories that an LLM can provide. A discussion with a teacher would include that competence but also, to different extents, some pressure to position oneself as competent in front of the teaching staff, whereas it is possible to bombard an LLM with questions without worrying about how the LLM is perceiving one as a student. It may be argued that there is still pressure to perform as competent in front of one's teammates, so an individual assignment may result in even more comfort when interacting with the LLM in this manner. 

The assignment sparked discussions and reflections -supported by the LLM- on critical thinking and action, including the equivalent for awareness and responsibility. Thus, it covered well an introduction to the ethical process by Lennerfors et al.~\cite{LennerforsFIE20}.

As the students point out, LLMs can provide different answers to the same (or similar) prompt due to, e.g.,  ambiguity in the text, sensitivity to the wording, and the randomness used during the text generation process~\cite{10.1145/3544548.3581388}. In particular, LLMs are probabilistic models that generate responses based on a probability distribution and choose one of the most likely ones~\cite{Sun_Shi_Gao_Ren_de_Rijke_Ren_2023}. Theoretically, an LLM could propose any option that an individual described as part of the training data. However, it is unlikely that the LLM will provide new reasonable alternatives unless they are hallucinated~\cite{Sun_Shi_Gao_Ren_de_Rijke_Ren_2023}. This supports the assignment format described in this report, where students first think individually and then share with their group (as in-class discussions and later for the writing), the discussion with an LLM being the second and not first part of the work.

Even though we could not find comparable experience reports in the context of computing ethics or focusing on the student perspective on an LLM as a discussion partner, other studies have shown students' positive expectations towards adopting LLMs and related tools as part of their coursework~\cite{Budhiraja_2024}. 
Due to the recent nature of LLMs and their rapid advancement, the majority of related work so far addresses LLMs' performance in solving (coding) problems~\cite{finnieansley2023my}, generating individual feedback~\cite{kiesler2023exploring}, building LLM-based assistants for conversational programming~\cite{qiconversational}, or course material for educators~\cite{Sarsa2022}. 
Prather et al.~\cite{prather2023its}, for example, explored student perceptions of the benefits and challenges of LLMs for learning, and design implications, but their focus was the novice programmer's perspective. The same applies to the study by MacNeil et al.~\cite{macneil2023experiences}. We expect the first comparable papers reporting on students' use of LLMs in computing course contexts other than programming to appear at SIGCSE 
and ITiCSE 2024.


\section{Limitations}

As it was a group assignment, we are not able to claim that the perspectives from all students were reflected in the submitted texts, as some students may have dominated the conversation more than others, or some may have participated less. While the in-class discussions were monitored, the interactions with LLMs were not.

The study is limited to the Swedish context, with students with no education experience in this country. Their English proficiency was diverse. For some, LLMs helped them express their thoughts, but there is a risk that others relied too heavily on ChatGPT.

\section{Conclusions and Future Work}

This experience report details the student perspectives on using an LLM as a discussion partner in computing education. Specifically, students analyzed an ethical dilemma using Lennerfors et al.'s ethical process \cite{LennerforsFIE20} and discussed the case with ChatGPT. This study differs from most previous work in that it is not focused on purely technical computing content, like writing code, and the emphasis is on the students' perspective as opposed to the educators' view.

The ethical dilemma for the assignment was connected to professional ethics, considering the impact of work-life practices on one's personal life and the planet. Future work can explore dilemmas connected to working with technology and working with others. In the latter, the biases of LLMs may be more apparent to students, or at the very least, a clearer opportunity for educators to bring up as part of their ethics module this important conversation connected to ethics and equal opportunities in computing.

Educators interested in designing assignments where LLMs are used as discussion partners may explore additional settings, e.g. individual discussions with the LLM that are then peer-reviewed.

\section*{Acknowledgements}
Johan Snider’s suggestions greatly inspired us.
We also thank Daniel Schiffner and the DIPF | Leibniz Institute for Research and Information in Education for their support.


\bibliographystyle{ACM-Reference-Format}
\balance
\bibliography{sample-base}

\end{document}